PLOS ONE

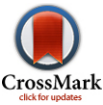

# Homophily and the Speed of Social Mobilization: The Effect of Acquired and Ascribed Traits

**Jeff Alstott[1,2], Stuart Madnick[3]\*, Chander Velu[4]**

1 Section on Critical Brain Dynamics, National Institute of Mental Health, Bethesda, Maryland, United States of America, 2 Behavioral and Clinical Neuroscience Institute, Departments of Experimental Psychology and Psychiatry, University of Cambridge, Cambridge, United Kingdom, 3 Information Technologies Group, Sloan School of Management and Engineering Systems Division, School of Engineering, Massachusetts Institute of Technology, Cambridge, Massachusetts, United States of America, 4 Institute for Manufacturing, Department of Engineering, University of Cambridge, Cambridge, United Kingdom

## Abstract

Large-scale mobilization of individuals across social networks is becoming increasingly prevalent in society. However, little is known about what affects the speed of social mobilization. Here we use a framed field experiment to identify and measure properties of individuals and their relationships that predict mobilization speed. We ran a global social mobilization contest and recorded personal traits of the participants and those they recruited. We studied the effects of ascribed traits (gender, age) and acquired traits (geography, and information source) on the speed of mobilization. We found that homophily, a preference for interacting with other individuals with similar traits, had a mixed role in social mobilization. Homophily was present for acquired traits, in which mobilization speed was faster when the recruiter and recruit had the same trait compared to different traits. In contrast, we did not find support for homophily for the ascribed traits. Instead, those traits had other, non-homophily effects: Females mobilized other females faster than males mobilized other males. Younger recruiters mobilized others faster, and older recruits mobilized slower. Recruits also mobilized faster when they first heard about the contest directly from the contest organization, and decreased in speed when hearing from less personal source types (e.g. family vs. media). These findings show that social mobilization includes dynamics that are unlike other, more passive forms of social activity propagation. These findings suggest relevant factors for engineering social mobilization tasks for increased speed.

**Funding:** The authors were partially funded by the NIH-Oxford-Cambridge Scholarship Program and by the John Norris Maguire Professorship in Information Technologies Chair account. Langley Castle Hotel funded the awards for the contest used in this study and gathered the data. Other than this, the funders had no role in the subsequent study design, data analysis, decision to publish, or preparation of the manuscript.

**Competing Interests:** Langley Castle Hotel sponsored the social mobilization competition that served as the basis for the data collected and analyzed. Co-author Stuart Madnick is one of the owners of Langley Castle Hotel. The data was analyzed after the competition was completed and, to the best of the authors' knowledge, Langley Castle Hotel gains no financial advantage as a result of the research reported. This does not alter the authors' adherence to all the PLOS ONE policies on sharing data and materials.

* E-mail: smadnick@mit.edu

## Introduction

Social mobilization is a movement to engage people's participation in achieving a specific goal through self-reliant efforts. Social mobilization can have a broad impact on society, as seen in social movements leading to change in culture or government policy [1–6]. Social mobilization also occurs on much smaller scales, such as friends spontaneously recruiting each other to run search and rescue operations [7,8]. The process of social mobilization can also be purposefully activated and directed; organizations and firms are increasingly turning to social mobilization and large-scale crowdsourcing to solve a variety of problems [9–11]. Using contests as a tool, initial principles have been identified for engineering social mobilization tasks to recruit large numbers of people for various purposes [12–14]. Although social mobilization has been studied extensively in a variety of contexts, there has been little attention to measuring the speed of mobilization and quantifying what personal traits predict that speed. We used a contest as a field experiment to measure how mobilization speed is affected by four traits: gender, age, geography, and information source. The homophily behaviors of these personal traits indicates that the dynamics of social mobilization are distinct from other forms of social activity propagation. These and other traits may be leveraged to engineer social mobilization tasks so as to recruit people more quickly.

Social activity typically spreads on an existing social network. Individuals can deliberately activate a single chain of social contacts to try to connect to a target person, as in the "six degrees of separation" scenario [15–17]. Social activations can also branch and spread without a target or purpose, such as with knowledge propagating through a network [18–21]. The behavior of these spreading social activations is in part governed by the traits of the individuals in the social network. Perhaps the best known effect is that of homophily, whereby activity is more likely to propagate between two individuals if they share a trait [22–24]. The literature differentiates between two types of homophily – status homophily and value homophily [22]. Status homophily includes sociodemographic dimensions that stratify society while value homophily includes internal states presumed to shape our orientation toward future behavior. We focus on status homophily in this study. Status homophily can be categorized as either ascribed or inherited traits, like gender or age, or acquired or achieved traits, such as geographical location [22]. Another potential driver of activation spreading is affinity, in which individuals are more likely to take up a particular activation (such as an idea or product) if they have some affinity for that particular subject [25]. Other agents who know the individual's affinities, then, may preferentially direct information to that individual that will be well received.

Analyses of social propagation phenomena have primarily described the length or extent of the activity spread, along with the factors that influence them. However, the speed of propagation can also be critical, particularly in time-sensitive domains like rescue operations or campaigning prior to elections. For various types of human communication the speed of activity propagation is heterogeneous and its distribution is heavy-tailed [26,27]. Demographic traits influencing speed have also been well characterized for such passive, diffusion-like processes as the spread of product adoption and musical tastes [28–30]. However, in the case of social mobilization, in which individuals are actively recruiting others for a purpose, our understanding of the predictors of speed of mobilization are still at a nascent stage.

Here we use a global social mobilization contest to study four personal traits and how they influence the speed of mobilization: gender, age, geography, and information source. Of these traits, our study shows that ascribed traits (gender, age) have no significant homophily effect on mobilization speed, whereas acquired traits (geography, information source) have significant homophily influence. Gender and age both have significant, non-homophily effects different from those reported in other contexts. Some types of information sources also yielded faster mobilization than others. These findings indicate that social mobilization speed has some elements in common with passive varieties of social activity propagation, but also has additional, distinct dynamics. A better understanding of these and other predictors of social mobilization speed may enable engineering of mobilization scenarios in order to achieve a particular objective rapidly.

## Results

We ran a global contest involving time-critical social mobilization, inspired by the Network Challenge contest organized by the Defense Advanced Research Projects Agency (DARPA) in the United States in 2009, which was won by the Red Balloon Challenge team by using a particular monetary incentive structure [12]. Our contest was for Langley Castle Hotel in Northumberland, United Kingdom. The task was to find five knights in parks throughout the United Kingdom on a particular weekend, each with an ID code written on their shield. Contest participants registered on the contest website, and could recruit other participants onto their team online in several ways (see Methods). Participants had financial incentive to form large teams by recruiting new members, who then recruited other members, and so on (example team structure, Fig. 1A). The first registered participant to correctly report the position of a knight was awarded £1,000. The discoverer's recruiter also received £500, the recruiter's recruiter received £250, and so on. This contest incentive structure was previously found to produce large social mobilization [12]. Any team that as a whole found more than one knight would also be awarded a £250 bonus, given to the team founder to distribute as desired. Additionally, the team founders of the first, second, and third largest teams received £1,000, £500, and £250 respectively.

Unlike the DARPA contest that was limited to a single country, two of these knights were "cyber knights", present not in the physical parks themselves but in Google Maps or Google Earth.

This allowed for participants outside of the United Kingdom to readily participate, and indeed over 30% of participants in the contest were from outside the UK.

## Team Creation and Dynamics

A total of 1,089 participants registered, with 148 starting their own team. Of the teams, 97 did not mobilize any other team members, leaving 51 teams that recruited new participants. Participants could act as both recruits (if they joined a team) and recruiters (if they mobilized others). In these teams, 152 participants acted as recruiters, mobilizing at least one other participant. These recruiters mobilized 941 recruits. The mean team size was 7.36, and the mean size of teams larger than 1 was 19.45.

To test the robustness of the observed dynamics of this social mobilization contest we compared the size and behavior of the teams to previously reported results from a contest using a similar incentive system [12]. This previous research had suggested the distributions of team size and of recruiters' number of recruits both followed power laws. Power laws are very heavy tailed probability distributions, and are notable because they imply the existence of extremely large events, such as a mobilization that grows to encompass the entire global social network. We examined the team dynamics in the present study using rigorous statistical methods [31,32], described in Methods, and found modest support for power laws. The parameter values of these power laws were consistent with those reported previously (Fig. 1B,C). This replication of previously described team dynamics indicates that at least some features of social mobilization are robust in this style of contest, in which participants recruit others into teams to find particular targets. We now extend the analysis of this type of contest to our primary focus, the speed at which new participants were recruited.

## Measuring and Modeling Mobilization Speed

When participants registered on the Langley Castle Hotel website to join a team we recorded their registration time. The time difference between when one participant registered and when a recruit they recruited registered was the speed of mobilization across that social connection, and is similar to the inter-signup time metric used in Pickard et al. [12]. The mean mobilization speed was 6.7 and the distribution was very heavy-tailed, with a standard deviation of 7.2 days (a histogram of mobilization speeds is shown in Fig. S1). There was one month between registration opening and the contest end date, and so the mobilization speed distribution was right-censored; the longest mobilization interval was 26.6 days. The key goal of this study was to understand the personal traits influencing these mobilization speeds.

We collected several pieces of information about the participants when they registered on the contest website (see Methods). We used this information to examine the influence of four traits on the speed of mobilization: gender, age, geography and the information source from which participants first heard about the contest (which could be a source other than their recruiter). We also controlled for other factors that could influence the speed of mobilization in order to account for their heterogeneity. We describe these control variables below. We modeled this speed of mobilization with a Cox proportional hazard model (see Methods), which has been used extensively to describe the spreading of epidemics and subsequently adopted to study diffusion processes on social networks, such as product adoption [28]. The Cox proportional hazard model measured the influence of the four main traits on the speed of mobilization, controlling for other relevant factors (for goodness of fit measures, see Information S1).

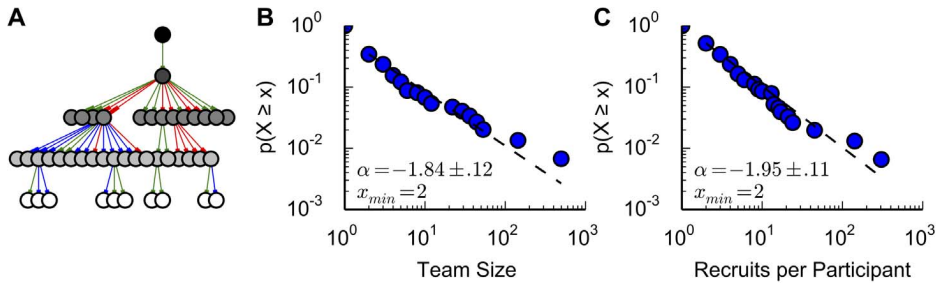

**Figure 1. Mobilized teams grew to a variety of sizes at a variety of rates.** (**A**) An example team growing from generations of recruiters to recruits, with different recruiter-recruit mobilizations having different types of links. The team starter's icon is black, and the future members decrease in shade as their generation in the team increases. Blue links indicate the recruiter and recruit heard about the contest through the same type of source (ex. friends). Red links indicate the recruiter and recruit heard through different types of sources (ex. family vs. the media). Green links indicate one or both participants did not give information on this personal trait. This example team was the 4th largest in the contest. (**B–C**) Using a similar social mobilization incentive system to that used in the present study, previous research suggested the distributions of team sizes and of recruiters' number of recruits followed power laws, with $\alpha$ of 1.96 and 1.69, respectively [12]. We used the statistical methods of Clauset et al. [31,32] to find weak to modest support for discrete power laws on these metrics, though the power laws' scaling parameters $\alpha$ are replicated. Distribution plots are complementary cumulative distributions (survival functions). (**B**) Team size. There were 148 teams, with 51 recruiting additional members beyond the founder. The power law fit was preferred over an exponential (LLR: 58.53, p<.01), but was no better of a fit than a lognormal (LLR:.01, p>.9) (**C**) Number of recruits for each recruiter. There were 1,089 participants, with 152 mobilizing at least one recruit. The power law fit was better than that of an exponential (LLR: 61.45, p<.02), but was not a stronger fit than the lognormal distribution (LLR:−.04, p>.9)
doi:10.1371/journal.pone.0095140.g001

A hazard function is the likelihood of an event occurring after some time $t$. In our hazard model, the hazard function at time $t$ was the likelihood of a recruit registering for the contest $t$ units of time after their recruiter had registered. The influence of a particular trait, such as geographic location, was observed by how much higher or lower the hazard was in the presence of that trait relative to a baseline. This increase or decrease in hazard to baseline was expressed as a hazard ratio. Higher hazard ratios reflected higher likelihoods of registering for the contest at all times $t$, which indicated a faster social mobilization speed. Lower hazard ratios, conversely, indicated slower social mobilization speed, through lower likelihoods of registering for all times, $t$.

The four personal traits can be classified as either ascribed or acquired traits [22]. Gender and age are ascribed traits. Geography and information source are acquired traits, as individuals can decide where to live or what information sources to pay attention to. Below we first discuss the effects of ascribed traits and then discuss acquired traits on recruitment speed. These findings are summarized in Table 1.

## Influence of Ascribed Traits: Gender and Age

**Influence of Gender.** A homophily effect was not supported in the case of gender, as mobilizations in which recruiter and recruit were the same gender were not significantly faster than different-gender mobilizations (p>.05). However, another effect was present: females mobilized other females faster than males mobilized other males (Fig. 2; p<.05). Recent research on the role of gender in the speed of product adoption spread has yielded conflicting findings on whether males or females have greater influence or susceptibility to influence [28,29]. In the present social mobilization task, the effect of influence was greatest when both recruiter and recruit were both female, and the least when the two were both male.

**Influence of Age.** Participants' ages were binned into 20-year ranges, and the proportional hazards model included the interaction of the recruit's age with the recruiter's age. A homophily effect was not supported in the case of age, as mobilization was not faster when the recruit and recruiter were of the same age group. However, the effect of the recruiter's and

**Table 1.** Summary of Findings.

| Personal Trait | Homophily Category | Homophily Effect Present | Findings |
|---|---|---|---|
| Gender | Ascribed | No | Mobilization was not significantly faster when the recruiter and recruit were the same gender, compared to different-gender mobilizations. However, females mobilized other females faster than males mobilized other males. |
| Age | Ascribed | No | Mobilization was not faster when the recruit and recruiter were of the same age group. However, for any given recruiter age group, mobilization speed increased with the recruit's age. For any given recruit age group, mobilization speed decreased with the recruiter's age. Therefore, young recruiters and old recruits displayed fast mobilization, while old recruiters and young recruits displayed slow mobilization. |
| Geography | Acquired | Yes | Mobilization speed was faster when the recruiter and recruit were in the same city, compared to when they were in different cities or countries |
| Information Source | Acquired | Yes | Mobilization speed was faster when both the recruiter and recruit first heard about the contest through the same type of source. Additionally, hearing about the contest from more intimate or psychologically close sources of information produced faster social mobilization. |

doi:10.1371/journal.pone.0095140.t001

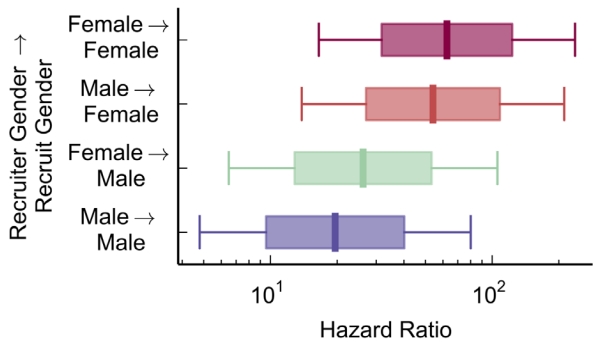

**Figure 2. Females mobilized other females faster than males mobilized other males.** No homophily effect was observed, as the recruiter and the recruit being of the same gender did not yield higher mobilization speeds. (p>.05). In all figures hazard ratios are the increase (>1) or decrease (<1) in likelihood of registering for the contest on a given day, reflecting an increase or decrease in mobilization speed. Boxes represent standard errors, and whiskers represent 95% confidence intervals. Redder boxes indicate faster mobilization (higher hazard ratios), while bluer boxes indicate slower mobilization (lower hazard ratios). Unless otherwise noted, the reference rate (hazard ratio = 1) is for participants who did not give data on that variable, or recruiter-recruit pairs in which at least one of the participants did not give data.
doi:10.1371/journal.pone.0095140.g002

recruit's ages on mobilization speed were still pronounced. For any given recruiter age group, mobilization speed increased with the recruit's age (Fig. 3A). This was in contrast to the main effect of recruit age (which did not include interaction with the recruiter age), which showed mobilization speed decreasing with recruit age. (Fig. 3B). Similarly, for any given recruit age group, mobilization speed decreased with the recruiter's age. (Fig. 3C, a rearrangement of the plots in Fig. 3A). Again, this was in contrast to the main effect of recruiter age, which showed mobilization speed increasing with recruiter age (Fig. 3D). These interactions of recruiter and recruit age are an instance of the Yule-Simpson paradox [33,34], in which two variables viewed in isolation appear to have one set of behaviors, but their interaction reveals that they in fact have an opposite set of behaviors.

The observed main effects of age mirror findings about age group's influence and susceptibility in passive product adoption diffusion [28]. However, when an interaction effect is present it supersedes that of any main effect, and we show that in social mobilization the role of age is reversed when interaction effects are considered. In particular, young recruiters and old recruits displayed fast mobilization, while old recruiters and young recruits displayed slow mobilization. A possible cause of this effect is that younger participants could motivate older participants, but older participants had difficulty motivating younger participants. This result of the interaction effect between recruits' and recruiters' ages contrasts with findings from passive product adoption diffusion, and shows no evidence of homophily influencing mobilization speed.

### Influence of Acquired Traits: Geography and Information Source

**Influence of Geography.** We find support for homophily in the case of geography, as social mobilization speed was faster when the recruiter and recruit were in the same city, compared to when they were in different cities or countries (Fig. 4; p<.01). This finding indicates that even in an era of increased telecommunications and "flattening" of the world, indeed even for this contest in which web registration was mandatory, geography is still important and influences how quickly teams mobilize.

**Influence of Initial Information Source.** Where the participant first heard about the contest, influenced mobilization speed. This could be a source other than their recruiter. As an example, a participant could first hear about the contest through a newspaper, then be recruited by another active participant who heard about the contest through a family member. In this case, the recruit and recruiter had heard about the contest from different information sources. In another case, it could be that the recruit and recruiter heard from the same type of information source. We find support for homophily in the case of information source, as mobilization was faster when both the recruiter and recruit first heard about the contest through the same type of source (Fig. 5A). Additionally, mobilization speed increased when the participant first heard about the contest directly from the Langley Castle organization (Fig. 5B). From the other categories of information source, the next highest speeds were from family members, then friends, down to the participant's organization or simply the media (difference of hazard ratios between "Langley Castle" and "Media", p<.01; all statistical tests on hazard ratio differences are derived from $\chi^2$ tests).

A direct communication between the participants and the organizers of the mobilization yielded the fastest mobilization speeds. In the absence of a direct exposure to the contest organization, however, this trend suggests that hearing about the contest from more intimate or psychologically close sources of information produced faster social mobilization.

### Control Variables

In order to isolate the effect of the personal traits we controlled for several other variables that could influence the speed of mobilization. We included these control variables in the Cox regression (see Methods). In recruitment contests such as this one the timing, generation and quantity of recruitments has previously been shown to influence mobilization [13,14,25,26], and therefore we control these factors by operationalizing them as described below. Every additional day after registration opened (meaning one less day until the contest began) the social mobilization speed increased, on average (Fig. S2, top), which is similar to deadline effects observed in other mobilization contests [14]. As teams grew, recruiters mobilized recruits, who then in turn became new recruiters mobilizing their own recruits. This process created "generations" of mobilization within a team. Each additional generation had slower mobilization relative to the one before it (Fig. S2, middle), similar to effects observed in the study by Rutherford et. al. [13]. Additionally, the more future recruits a participant would have, the faster that participant mobilized (Fig. S2, bottom). While causality obviously does not allow a participant's number of future recruits to directly affect his or her own mobilization speed, the statistical relationship indicates that those who mobilized quickly also recruited more recruits, independent of other factors.

### Discussion

As social mobilization becomes increasingly prevalent, the ability to engineer and influence the dynamics of mobilization will become ever more important within society. We replicated a contest designed to mobilize a large number of people, finding similar statistics of team size and growth to those reported in previous studies. We measured participants' mobilization speed and what personal traits were associated with the speed of social mobilization. We found that homophily on acquired traits

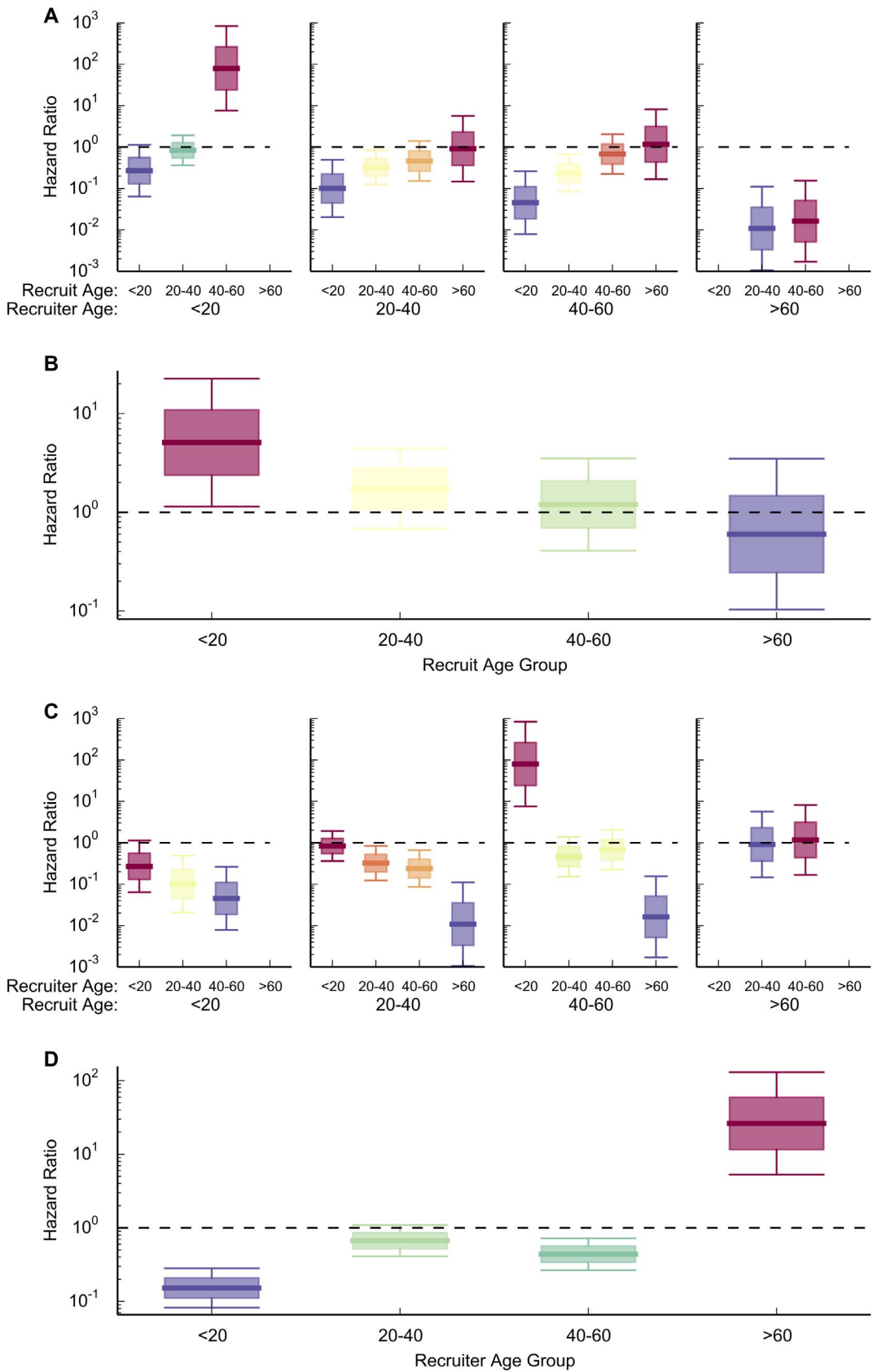

**Figure 3. Older recruits and younger recruiters had faster mobilization speeds, as revealed by the interaction of recruiter and recruit age.** In the Yule-Simpson paradox the interaction effect of two factors contrasts with the main effect of either factor taken individually, as is the case with recruit and recruiter ages' relationship with mobilization speed. In such a case the interaction effect supersedes the main effect. Absent

plots indicate no data for that interaction. (**A**) The interaction of recruiter and recruit age group on mobilization time, grouped by the recruiter's age. For any given recruiter age group, mobilization speed increased with the recruit's age. (**B**) The main effect of the recruit's age group on mobilization speed, which had the opposite behavior of that found in the interaction effect seen in (A). (**C**) The interaction of recruiter and recruit age group on mobilization time, grouped by the recruit age. For any given recruit age group, mobilization speed decreased with the recruiter's age. This is a simple rearrangement of the information in (A). (**D**) The main effect of the recruiter's age group on mobilization speed, which has the opposite behavior of that found in the interaction effect seen in (B).
doi:10.1371/journal.pone.0095140.g003

(geography and information source used) increased mobilization speed, while homophily was not present on ascribed traits (gender and age). Additionally, mobilization speed was faster when recruits heard about the contest from more personal sources. Gender and age, while not displaying homophily effects, were also found to have different influences on active social mobilization than those reported in more passive social activity propagation: Females mobilized other females more quickly than males mobilized other males; younger people mobilized others quickly while older recruits were mobilized more quickly.

The present findings provide a preliminary quantitative understanding that mobilization speed is a function of readily measurable personal traits. Furthermore, the influence of these traits is not necessarily the same as in other social activity propagation contexts. Homophily of ascribed traits, for example, has been previously shown to be very influential in passive, diffusion-like activity spreading, but in active mobilizations we did not observe any homophily effect for such traits. Age's role in social mobilization is also opposite to that observed in product adoption influence. In the active mobilization, younger individuals mobilize others faster and older individuals are mobilized more quickly. In passive influencing contexts, influence increases with age and susceptibility to influence decreases with age [28]. However, there are dimensions where social mobilization has similar dynamics to other forms of social activity propagation: acquired traits have a significant homophily effect. Additionally, hearing about the contest from a psychologically closer source may be due to those sources being similar to the recruit, coupled with a homophily effect. However, it could also be the case that the closer information sources know the recruit's preferences and have notified the recruit of the contest because they think the recruit has an affinity for the topic [25]. Such affinity would then increase the speed of the mobilization.

This contest was a framed field experiment, using a voluntary, non-randomized pool of subjects and natural field conditions in many elements of the experiment [35]. Individuals self-selected to participate by joining teams for a specific kind of contest, which involved finding knights for prize money. We controlled for factors that were observed and recorded. It could be that some observed effects (e.g. gender differences) are actually due to other, unmeasured factors (e.g. employment levels). It could also be that voluntary participation created a selection bias for participants who were attracted to this type of contest, who may behave differently from others (these issues are discussed further in Methods). Future studies can improve on these issues by reproducing the results in other contexts and controlling for additional factors.

Large-scale social mobilizations are becoming increasingly common and relevant, and often the speed of recruitment is critical to their success. A disease prevention campaign, for example, may need to propagate best practices against a new virus quickly. After a natural disaster, donation networks that are set up quickly could provide funds immediately. For those organizing such mobilization tasks, a greater understanding of the personal traits driving mobilization speed could improve the odds of success. By engineering a few elements of a mobilization task, it could be possible to increase the speed of recruitment. The predictors of social mobilization speed described here compose an initial set of possibly relevant personal traits, and opens the door for identification of additional factors and further research.

## Methods

We ran a large-scale social mobilization contest as a 'framed field experiment' [35], in which subjects were able to join the experiment and use their experience and knowledge in their unaltered natural field setting in making decisions. The contest was advertised by Langley Castle, through its web site www.langleycastle.com, newsletters, Facebook pages, email lists, and press releases. A copy of the master press release can be found at [36] and an amusing video is at [37]. Participants registered on the contest website, where they were directed to give demographic information about themselves and how they heard about the contest. Participants could register with their email address and making a password on the site, or alternatively through Facebook Connect. Registration on the website opened June 1, 2011. The competition started on July 2, 2011 at 9am and ended at 9pm on July 3, 2011. The "real" knights were in their assigned parks from roughly 9am to 9pm each day. The "cyber" knights were present on Google Maps and Google Earth all day.

Participants who registered using Facebook Connect could, at the end of the registration process, invite their Facebook friends to join the contest under their team. Registered participants were also provided with a URL they could share with others to register through, which would automatically put those new participants on their recruiter's team. In addition to the URL, the participants were also given a number that other participants could enter manually to register as their recruit.

Participants submitted information on knights they found through a form on the website, which required the inclusion of a code unique to each knight printed on their shield. Knights that were already found were announced on the contest website.

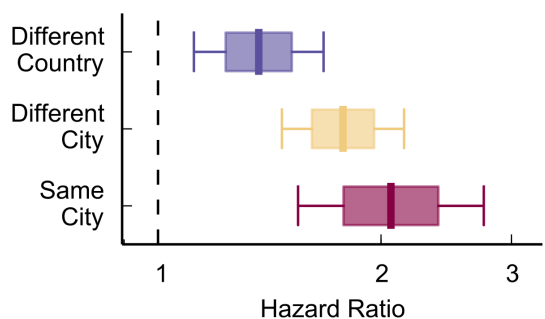

**Figure 4. Geographically closer relationships had faster mobilization speed.** Social mobilization was faster when the recruiter and recruit are in the same city, and slowest when they were in different countries.
doi:10.1371/journal.pone.0095140.g004

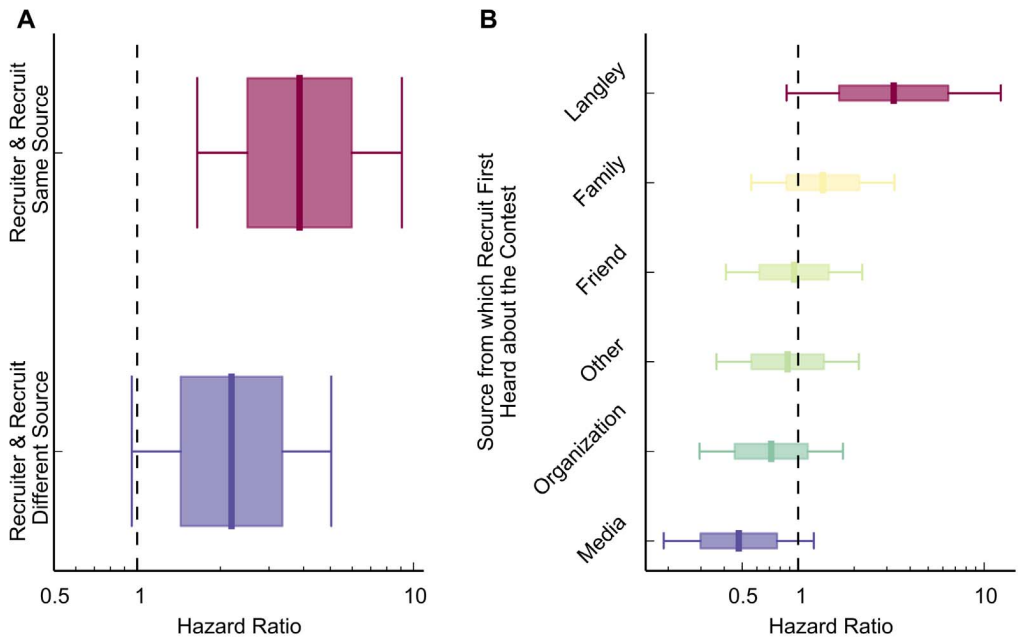

**Figure 5. Where a participant heard about the contest affected mobilization speed. (A)** Mobilization speed was fastest when participants hear about the contest from Langley Castle directly, and decreased as the source is more psychologically distant from the participant. **(B)** Mobilization was faster when the recruiter and recruit heard about the contest through the same category of source than when they heard through different sources.
doi:10.1371/journal.pone.0095140.g005

## Personal Information Collected

The participants' geographic location were inferred from the IP addresses they used to communicate with the contest website. Participants whose IP addresses could not be localized to the city level were treated as not having geographic data. The participant registration form's personal information questions and response categories were:

"Heard about us from": Friend, Family member, Your organization, Langley Castle, Media, Other

"Gender": Male, Female

"Age range": <20 years old, 20–40 years old, 40–60 years old, >60 years old

Participants were not required to respond on any question. The number of recruits with available data on a category and the number of recruiter-recruit pairings in which both participants had available data on a category were:

"Heard about us from": 505, 426

"Gender": 774, 756

"Age range": 529, 475

Geographic information: 704, 637

To safeguard personal privacy, participants' data was anonymized before analysis by removal of names, email address, and IP addresses. In addition, every contestant was required to acknowledge the Terms and Conditions of the competition, which, among other things, require that: "The competition is only open to persons who are 18 years of age or older at the time of entry." This anonymized dataset is available by contacting the authors.

## Proportional Hazards Model of Mobilization Speeds

We modeled the speed of mobilization using a Cox proportional hazard model, which models the time for an event to occur [38]. The time modeled was the interval between when a recruiter and recruit registered, with the time interval beginning when the recruiter registered and ending when the recruit registered. The probability of registration occurring (the "hazard") changes over time, creating a hazard function. The modeled hazard function is the likelihood of a recruit registering a given time interval after the recruiter has registered. The hazard function for a particular factor value (e.g. "male") is compared to a hazard function of another, reference factor value (e.g. "female"). The influence of the particular factor value is expressed as the ratio of these two hazard functions: the hazard ratio. Higher hazard ratios reflect higher likelihoods to register at any given time, which when multiplied successively across time speed up the point at which registration will occur. Thus, high hazard ratios indicate fast mobilization speed. Conversely, low hazard ratios extend the time until registration will occur, reflecting slower mobilization speed.

The personal traits included as factors in the model were:

- the recruiter's and recruit's ages and their interaction
- the recruiter's and recruit's genders and their interaction
- the type of source from which the recruit heard about the contest
- if the recruiter and recruit heard about the contest from the same type of source
- if the recruiter and recruit were in different countries, different cities in the same country, or the same city

The control variables, also included as factors to account for heterogeneity they were associated with, were:

- the number of recruits the recruiter had
- the number of recruits the recruit would have
- the generation the recruit was in the team
- the time since registration opened (the inverse of the time remaining until the contest began), which was expressed as the date the recruiter registered

With the recruit's factors described in a set $X_c$, and the recruiter's factors described in a set $X_p$, and the recruit's and recruiter's ages and genders represented in the set $S(X_c, X_p)$, our model had the form:

$$\lambda_c(t, X_c, X_p) = \lambda_0(t)\exp[X_c\beta^c + X_p\beta^p + S(X_c, X_p)\beta^{c \cdot p}]$$

With $\lambda_0$ as the baseline hazard function as a function of time since the recruiter's registration $t$, and $\lambda_c(t, X_c, X_p)$ as the hazard for a recruit at time $t$ with factors $X_c$ and whose recruiter had factors $X_p$. The coefficient $\beta^c$ is the effect of the recruit's factors on the hazard, $\beta^p$ is the effect of the recruiter's factors, and $\beta^{c \cdot p}$ the effect of their interaction, for certain factors (age and gender).

### Power Law Tests of Team Dynamics

We used the statistical methods of [31], as implemented in the powerlaw Python package [32],to evaluate whether several features of team dynamics were well-described by power law distributions. These features were the distributions of the number of generations in a team, team size, and a recruiter's number of recruits. The statistical methods included using a loglikelihood ratio (LLR) test between a best-fit power law (found through maximum likelihood methods) and an alternative distribution. A positive LLR indicates the power law fit is more likely, and a negative shows the alternative distribution is more likely. The significance of that LLR, however, is given by a p-value. A statistically insignificant LLR means the data does not clearly fit either of the candidate distributions more than the other. Lastly, the best-fit power law may not cover the entire distribution, but only be a good fit beyond a certain value, the $x_{min}$. The shape of these distributions does not impact the use of the Cox proportional hazards model for describing mobilization speed.

### Advantages and Disadvantages of Framed Field Experiment Methodology

There could be two major concerns regarding our field experiment methodology: sample selection and unobserved factors.

**Sample selection.** This framed field experiment uses a voluntary non-randomized subject pool, which are typically done as close to the real environment as possible with minimum alterations to the context to avoid influencing subject behavior and other biases that might be due to the design of the data collection. Since the pool of subjects joined the contest voluntarily without us administering any process of randomization, there might be a self-selection bias in that people attracted to the structure and themes of this contest may behave differently from those not attracted to them.

**Unobserved factors.** We measured various factors that could influence mobilization speed, including gender, age, geography and information source. We controlled for other factors, such as timing, generation and quantity of recruitments, but were limited to those factors that were observed and recorded. This leaves the possibility that other factors influenced the observations. For example, it could be that males and females had different employment rates, and it was this factor that led to their differences in social mobilization behavior. There are countless such possible confounding factors (such as females might have more time available, are harder workers, are smarter, etc.), some of which are even unobservable, making perfect measurement an impossible task.

**Mitigations.** Several major studies of social mobilization and other forms of social influence are also framed field experiments (e.g. [12,28]). Such studies have had similar limitations of sample selection and number of factors observed. In order to mitigate these limitations, rigorous methods have been developed for data collection and analysis. We use these methods, with all limitations acknowledged, to begin to identify how personal traits affect the speed of social mobilization. Quantitative studies of social mobilization speed are rare, and to the best of our knowledge the key studies in this area make no effort to measure most of the traits that we examine. By measuring factors that predict social mobilization speed, this work advances our understanding of this important phenomenon.

## Supporting Information

**Figure S1** The distribution of mobilization speeds was heavy-tailed. Mobilization speeds were measured by the interval between when a recruiter registered on the contest website and when their recruit registered. The mean mobilization speed was 6.7 days, with a standard deviation of 7.2 days.
(TIFF)

**Figure S2** Time left in the contest, additional generations, and additional future recruits all affected mobilization speed. The further in time the recruiting happened (i.e. closer to the contest date), the faster the mobilization speed. In contrast, as a team grew with generations of recruiters recruiting recruits, each additional generation beyond the first (hazard ratio = 1) slowed down mobilization speed. The recruit's mobilization speed increased for each additional future recruit he or she had beyond zero.
(TIFF)

**Information S1** Goodness of fit measures for the Cox proportional hazards model.
(PDF)

**Code S1** Anonymized data and code used to produce the reported analyses.
(ZIP)

## Acknowledgments

The authors would like to thank Anton Phillips for operational support as the general manager at Langley Castle, Sunny Cheung for designing and implementing the web site software, and Wei Pan for insights and suggestions based on his experience with the DARPA Red Balloons Challenge experiment.

## Author Contributions

Conceived and designed the experiments: SM CV. Performed the experiments: SM CV. Analyzed the data: JA. Wrote the paper: JA SM CV.